# Displacive Excitation of Coherent Phonons in YBa$_2$Cu$_3$O$_7$: An Optical Evidence for the *Van-Hove* Singularity

W.-S. Zeng and J. Kuhl

*Max-Planck-Institut für Festkörperforschung, Heisenbergstrasse 1, D-70569 Stuttgart, Germany*

(August 13, 1996)

The amplitude of displacive excitation of coherent phonons (*DECP*) in YBa$_2$Cu$_3$O$_7$ observed in time-resolved spectroscopy increases by a very large factor of 15 between 20 K (below $T_c$) and 100 K (above $T_c$). Microscopic modelling of phonon excitation reveals that these amplitude changes of the $A_{1g}$ mode phonons are related to the existence of the *Van-Hove* singularity (VHS) below but very close to Fermi energy.

Femtosecond pump-probe (*PP*) measurements in YBa$_2$Cu$_3$O$_{7-\delta}$ indicated that ultrafast optical excitation causes oscillations in the photo-induced reflectivity[1,2]. For YBa$_2$Cu$_3$O$_7$, the oscillation frequencies determined by fast Fourier transformation (FFT) match with two Raman active $A_{1g}$ modes, corresponding to vibrations of Cu(2) (150 cm$^{-1}$) and Ba (120 cm$^{-1}$)[1]. Among all experimental facts, the strong growth of the oscillation amplitude between 20 and 100 K is most striking. Between 20 and 100 K, the intensities of the 150 cm$^{-1}$ and the 120 cm$^{-1}$ mode change by a factor of 2 and 15, respectively. Mazin *et al.*[3] interpreted this experimental evidence by a phenomenological model[4] with some local density approximation (LDA) parameters and within the BCS framework of an isotropic gap. The different change of relative intensities of the Cu(2) and the Ba mode below and above T$_c$ was interpreted as a consequence of the mixing of these two modes. The mechanisms of *DECP* above and below T$_c$ are phenomenologically different. Above T$_c$, the driving force for *DECP* is the usual interband transition and thus mixing of the two modes contributes little to the observed *DECP*. Below T$_c$, however, the *DECP* is additionally excited by the movement of the ions to their normal equilibrium positions caused by the destruction of the superconducting pairs after absorption of the pump laser. In this case, the mixing of these two modes becomes much more important and results in a remarkable reduction of the Cu(2) mode. Although this phenomenological model captures the essential physics in the *PP* process (e.g, this model assumes the driving force for the *DECP* is proportional to the density of excited electrons), it does not consider the real physics on the microscopic level.

In this Letter, we analyze the given experimental facts on a microscopic level combining the microscopic theory for *DECP* proposed by Kuznetsov *et al.*[5] with the eight-band Hamiltonian model developed by Andersen *et al*[6]. Our results indicate that the mechanism of *DECP* below and above T$_c$ is microscopically different as claimed previously[3] and that the strong enhancement of the amplitude of *DECP* below T$_c$ is additionally caused by the existence of the *Van-Hove* singularity (VHS) approximately 20 meV below the Fermi energy. We also study the influence of the gap symmetry on the amplitude of *DECP*. According to Ref.[5], the amplitude of the *DECP* for an $A_{1g}$ mode should satisfy the following equation of motion:

$$\frac{\partial^2}{\partial t^2} D_{A_{1g}} + \omega^2_{A_{1g}} D_{A_{1g}} = -2\omega_{A_{1g}} \sum_{\mathbf{k}} M_{\mathbf{k}} <c^{\dagger}_{\mathbf{k}} c_{\mathbf{k}}> . \quad (1)$$

Here $M_{\mathbf{k}}$ and $<c^{\dagger}_{\mathbf{k}} c_{\mathbf{k}}>$ are the electron-phonon (EP) coupling and excited electronic density matrix in **k** space, respectively.

It is known that three bands cross the Fermi energy (E$_F$), two for electrons in the CuO$_2$ plane and one of the CuO linear chain for the YBa$_2$Cu$_3$O$_7$. Many important physical properties are due to the energy band near E$_F$. The eight-band model[6] is an effective, nearest neighbour, tight-binding model Hamiltonian for a bilayer CuO$_2$ and describes very well the two "plane-bands" crossing E$_F$. The Hamiltonian in Ref. [6] includes the four $\sigma$ orbitals: $\psi^s_1(Cud_{x^2-y^2})$, $\psi^s_2(Cus)$, $\psi^s_3(O2p_x)$, $\psi^s_4(O3p_y)$ and four $\pi$ orbitals: $\psi^s_5(O2p_z)$, $\psi^s_6(O3p_z)$, $\psi^s_7(Cud_{zx})$, $\psi^s_8(Cud_{zy})$. This reduced Hamiltonian is a direct result of the so-called down-folding procedure. It is interesting to note that this model can describe the form of the extended saddle-points around the Fermi energy in the Brillouin zone (BZ). The corresponding VHS in the density of states (DOS) is regarded as the reason for enhancement of the superconductivity transition temperature[7]. Next, we will show in Fig.1 the calculated the EP coupling matrix in the frame of the above bilayer eight-band tight-binding(TB) model. A more general description of calculating EP can be found in Refs.[8,9]. After the standard transformation[10] into the momentum representation and diagonalization of the TB Hamiltonian of eight-bands in Ref. [6], we obtain the reduced Hamiltonian including the EP coupling and get the EP coupling matrix for the $A_{1g}$ mode[8]:

$$M_{A_{1g}}(\mathbf{k}, \mathbf{q}=0) = eE_z \sqrt{\frac{\hbar}{2M_O \omega_{A_{1g}}}} \frac{1}{\sqrt{2+M_{Cu}/M_O}}$$
$$\times [\sum_{s,j=3,4,5,6} \mu^2_{s,j}(7,\mathbf{k})) - (\mathbf{E}_b)/E_z \sum_{s,j=1,2,7,8} \mu^2_{s,j}(7,\mathbf{k})] \quad (2)$$

where $\mu_{s,j}(7,\mathbf{k})$ are the coefficients of the diagonalizing the $H^8(s)$ Hamiltonian contributed by different atomic orbitals (j), bottom (s=0) and top (s=1) layer, respectively. Other notations are same as in Refs.[8,9].

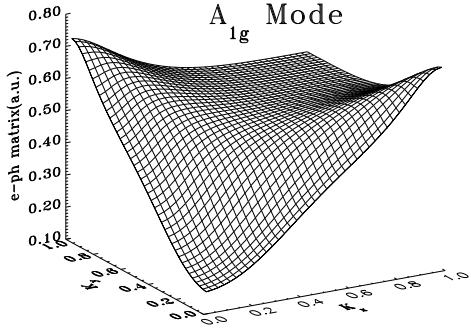

FIG 1. The evaluated EP coupling matrix $M_{A_{1g}}(\mathbf{k})$ for the $A_{1g}$ mode of the top layer (s=1) in the bilayer $CuO_2$ system, as a function of $\mathbf{k}$ in the 2D lattice. The parameters for calculating Eq.(2) are taken from the diagonalization of the Hamiltonian in Ref.[6].

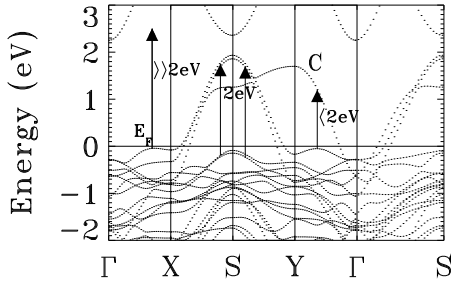

FIG 2. The LDA band structure of $YBa_2Cu_3O_7$ (from Ref.[6]). For 2 eV photoexcitation, the inter-plane-band transition from saddle-points around point $\mathbf{X}$ or $\mathbf{Y}$ are forbidden in the normal state because the transition energy is much larger than 2 eV.

The band structure of $YBa_2Cu_3O_7$ in the normal state obtained from LDA calculations is shown in Fig.2. The saddle points corresponding to the VHS are predicted around the $\mathbf{X}$ and $\mathbf{Y}$ points as confirmed by photoemission results[11]. It can be seen, at normal state, that the optical inter-plane-band transitions (shown by the arrows in Fig.2) around $\mathbf{X}$ or $\mathbf{Y}$ are impossible for 2eV-photoexcitation because the transition energy is much higher than 2 eV. Another direct optical transition would be possible from the saddle-point around $\mathbf{Y}$ to the chain band C. This transition plays an important role for the relaxation of electrons and shows substantial changes of the $PP$ signal for $YBa_2Cu_3O_{7-\delta}$ samples with different oxygen content[12], but it has no effect on the excitation of the phonons in $CuO_2$ plane. We will not consider this transition in this paper. Thus, only electrons in the vicinity of the $\mathbf{S}$-point can be excited and the corresponding integration conditions for the right side of Eq(1) would be: ( where $\eta$ can be choosen as $\pm\pi/2a$ )

( a.): $\mathbf{k} \in (\mathbf{k_S} + \eta)$.

This result is strongly supported by the density of states(DOS). Figure 3 shows the total (DOS) calculated from the bilayer eight-band TB model. The DOS exhibits two *Van-Hove* singularities (VHSs) below $E_F$: one very close to $E_F$ ( approximately 20 meV below ) and another much farther away. In the eight-band TB model, the wave functions are divided into *odd* and *even* with respect to the mirror plane through the yttrium atom in the unit cell of $YBa_2Cu_3O_7$. The corresponding VHS for an *even* band is indicated by the solid line and the DOS for an *odd* band by the dashed line. Further calculations indicate that the main contribution to the DOS around the $\mathbf{S}$ point near the Fermi energy $E_F$ comes from odd rather than from even bands. Therefore, only electrons described by the dashed line will thus contribute to the $DECP$ in the normal state.

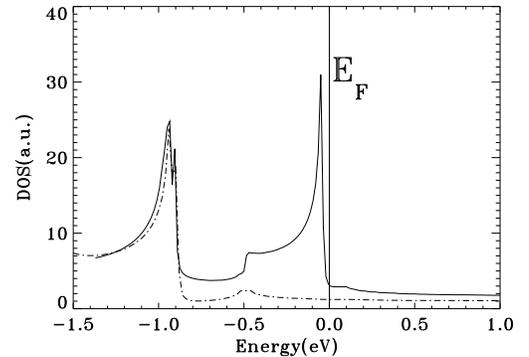

FIG 3. The total density of states (DOS) calculated from the eight-band TB model (solid line). The dashed line indicates the contribution of the **even**-bands ( see text ).

The interband transition ( i.e. electron/photon interaction alone) hardly can explain[3] the substantial change of the reflectivity below $T_c$. In the superconducting state, however, breaking of Cooper pairs which may be also accomplished via electron/electron or electron/phonon interaction represents an alternative mechanism for $DECP$. Although direct interband photoexcitation of unpaired electrons and Cooper pairs at the saddle points around $\mathbf{X}$ or $\mathbf{Y}$ is still impossible, effective Cooper pair breaking can occur via collisions with unpaired carriers excited by the 2 eV photons at the $\mathbf{S}$ point. Additionally the fast relaxation of the photo-excited paired- and unpaired-electrons around the **S** point in the BZ results in the generation of a large density of phonons whose energy is comparable or larger than the superconducting gap. Breaking of Cooper pairs at the **X** and **Y** point via electron/electron or electron/phonon scattering can explain the drastic increase of the the driving force for *DECP* in the superconducting state because of the existence of the VHS at these points in the BZ. As shown in Fig.3, the VHS around $E_F$ comes from the *even* band. The excited electrons will be described by the solid line in Fig.3 in the superconducting state. The limits for the integration of the right side of Eq.(1) in the BZ are now given by

(b): $\mathbf{k} \in \mathbf{BZ}$

Next we will try to estimate the *DECP* in the superconducting state. For simplicity let us assume at the beginning that the gap is isotropic. Then the excitation energy of quasiparticles in the superconducting state is[13]

$$\xi_{\mathbf{k},s} = \sqrt{\epsilon_{\mathbf{k},s}^2 + \Delta(\mathbf{k})^2} \qquad (3)$$

The actual number of broken pairs in general would be less than the total number of superconducting pairs. However, the LDA estimation in Ref.[3] shows that the actual number of broken pairs will be very close the total number of superconducting pairs in the ground state for the typical excitation intensities applied in PP experiments. Therefore we can take $\sum_{\mathbf{k},s} \delta(\xi_{\mathbf{k},s} - E)$ to calculate the number of excited electrons and their contribution to the driving force for *DECP* in the superconducting state. In addition to pair breaking, the pump pulse also excites unpaired electrons whose contribution is assumed to be identical with the normal state. Thus according to Eq.(1) and the integration limits $a$ and $b$, the ratio of the amplitudes (ROA) of *DECP* below and above $T_c$ will be

$$\frac{D_{A_{1g},s}}{D_{A_{1g},n}} = \frac{\sum_{s,\mathbf{k}\in\mathbf{BZ}} M_{s,\mathbf{k}}^7 \delta(\xi_{\mathbf{k},s} - E_F) + \sum_{\mathbf{k}\in(\mathbf{k}_S \pm \eta)} M_{1,\mathbf{k}}^7 \delta(\epsilon_{\mathbf{k},1} - E_F)}{\sum_{s=1,\mathbf{k}\in(\mathbf{k}_S \pm \eta)} M_{s,\mathbf{k}}^7 \delta(\epsilon_{\mathbf{k},s} - E_F)} \qquad (4)$$

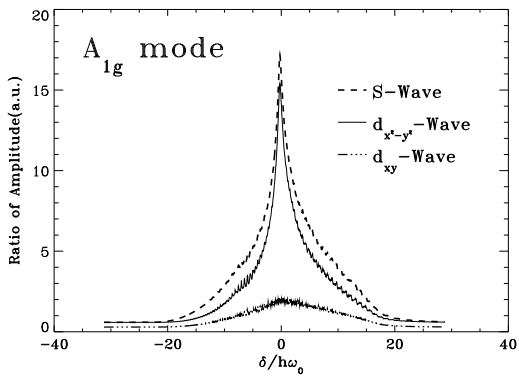

FIG 4. Numerically calculated ratio $D_{A_{1g},s}/D_{A_{1g},n}$ (see Eq.(4)) as a function of the Fermi energy shift $\delta$ normalized to the energy of $A_{1g}$ mode. For the d-wave case $d_{x^2-y^2}$ and $d_{xy}$ gap symmetry are considered.

The Van-Hove scenario will easily explain the occurrence of a high superconducting transition temperature for several pairing mechanisms[14]. Indirect evidence for the existence of such a singularity has been reported in studies of the specific heat, thermopower, isotope effect, *etc.* Recently, direct evidence for a saddle-point, related to the $CuO_2$ planes and located about 19 meV below the Fermi energy $E_F$ at the **Y** or **X** point in the BZ, has been obtained in high energy resolution, angle-resolved photoemission spectroscopy (ARPES) on $YBa_2Cu_4O_8$[11]. It is worthwhile to note, however, that optical studies in which the rather high **k**-sensitivity should be assumed have not led to any signature for the VHS in the cuprates so far. In the following, we will demonstrate that the 15 times increased amplitude of the $A_{1g}$ mode yields additional evidence for the existence of the VHS approximately 20 meV below the Fermi energy. In the normal state, as we have discussed before, the driving force for the *DECP* is the 2 eV interband transition optically allowed only around the **S** point in BZ as shown in Fig.2. Thus, the total number of excited electrons will be very small as shown in Fig.3. In the superconducting state, however, in addition to the interband transitions, effective pair breaking will contribute to the driving force on the right side of Eq.(1). More important, the contribution from pair breaking will become much larger if there exists a VHS below but close to $E_F$. Consequently, the huge number of electrons located around the VHS will be involved in the *PP* process in the superconducting state. The ratio of amplitudes (ROA) of DECP below and above $T_c$ can be calculated as a function of the shift $\delta$ of the Fermi energy normalized to the phonon energy of the $A_{1g}$ mode. This can be done by replacing $E_F$ in Eq.(4) by $E_F + \delta$. Positive shift means that the VHS is farther away from $E_F$. Figure 4 shows the sensibility of the VHS position on the ROA. It can be noted that the calculated results agree with PP measurement quite well in the case that the VHS is around the Fermi level. Consequently, it is impossible to observe an increase of the DECP amplitude below $T_c$ by a factor of approximately 15 if the VHS does not exist or is far away from the Fermi energy. Although Dagotto *et al* proposed explanation of the flat bands as coming from the correlation effects rather than band structure[15], the origin of the flat bands is still unclear. In this Letter, we showed that the VHS described well by one-electron approximate LMTO calculation can explain quantitatively the changes of DECP of $A_{1g}$ phonon in the PP measurements. Actually, as shown by Eq.(1), VHS scenario, no matter whether it comes from strong correlation effects or band structure, can explain qualitatively the optical data in the time-resolved spectra.

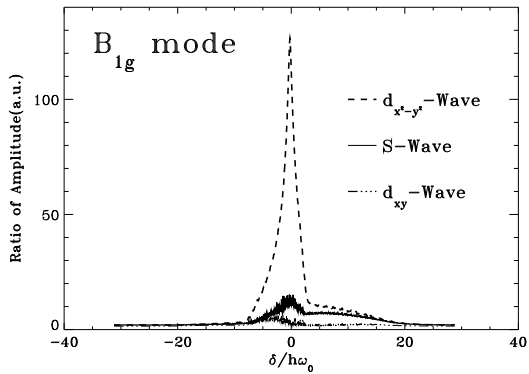

FIG 5. Numerically calculated results of equation (4) as a function of the Fermi-energy shift $\delta$ normalized to the phonon energy $\hbar\omega_0$ of the $B_{1g}$ for either s or d-wave ( with $d_{x^2-y^2}$ and $d_{xy}$ gap symmetry) coupling.

Finally the influence of the gap symmetry (GS) is now considered. This can be done by simply inserting the anisotropic gap equation $\Delta(\mathbf{k})=\Delta_0\zeta(\mathbf{k})$ into Eqs.(3) and (4). GS is $d_{x^2-y^2}$ with $\zeta(\mathbf{k}) = (cosk_x - cosk_y)/2$ and $d_{xy}$ when $\zeta(\mathbf{k}) = sink_x sink_y$. For $d_{xy}$ GS, the nodes will appear at the **X** and **Y** points (i.e. around the saddle-points of the VHS). In this case the carriers at the VHS form no Cooper pairs and thus the VHS cannot contribute to the driving force for the *DECP*. Therefore the result of the *PP* experiment excludes $d_{xy}$ GS. The numerically calculated curve shown in Fig.4 indicates that the ROA of the $A_{1g}$ mode increases by a factor of 16 for $d_{x^2-y^2}$ pairing, which is almost same as the value of 17 predicted for an isotropic gap. However, for the $B_{1g}$ or $B_{2g}$ mode, the ROA of *DECP* is even much more sensitive to the symmetry of the gap. In Fig.5, we show the same calculations for the $B_{1g}$ mode. The ROA of this mode will increase by a factor of 110 for $d_{x^2-y^2}$ pairing, as compared to a factor of 20 for the isotropic pairing. Again, ROA for $d_{xy}$ is also much smaller than value for other gap symmetry. Unfortunately, no experimental data have been reported for the *DECP* of the $B_{1g}$ or $B_{2g}$ mode so far. We would like to emphasize, however, that according to our theoretical model studies of $B_{1g}$ or $B_{2g}$ modes in optical experiments may provide a powerful tool to discern the gap symmetry.

In summary, we have shown that the VHS scenario can lead to very large amplitude of the *DECP* for $YBa_2Cu_3O_7$ below $T_c$. The mechanisms for excitation of the *DECP* below and above $T_c$ are microscopically different. The influence of the gap symmetry on the amplitude of the $A_{1g}$ mode is rather small but becomes very large for $B_{1g}$ or $B_{2g}$ modes. The model calculations suggest that for $A_{1g}$ mode, ROA$\sim$15, while for $B_{1g}$ mode, ROA is predicted as $\sim$110 for $d_{x^2-y^2}$-wave and $\sim$20 for s-wave symmetry.

W.-S. Zeng would like to acknowledge financial support by the Max-Planck-Gesellschaft during his stay at the Max-Planck-Institut für Festkörperforschung, Stuttgart. The encouragement and great help from O.K.Andersen are highly appreciated. We are grateful to A.I. Liechtenstein for valuable discussions on the EP matrix calculations and critical reading of the manuscript. We wish to thank R.Tank for technical help in preparing the figures for this work.

[1] W. Albrecht, Th. Kruse, and H. Kurz, Phys. Rev. Lett. **69**,1451 (1992).
[2] W. Albrecht, Th. Kruse, and H. Kurz, Appl. Phys. **A56**, 463 (1993).
[3] I.I. Mazin, A.I. Liechtenstein, O. Jepsen, O.K. Andersen, and C.O. Rodriguez, Phys. Rev. **B49**, 9210 (1994).
[4] H.J. Zeiger, J. Vidal, T.K. Cheng, E.P. Ippen, G. Dresselhaus and M.S. Dresselhaus, Phys. Rev. **B45**, 768 (1992).
[5] A.V. Kuznetsov and C.J. Stanton, Phys. Rev. Lett. **73**, 3243 (1994).
[6] O.K. Andersen, O. Jepsen, A.I. Liechtenstein and I.I. Mazin, Phys. Rev. **B49**, 4145 (1994) ; O.K. Andersen, A.I. Liechtenstein, O. Jepsen, and F. Paulsen, J.Phys.Chem.Solid **56**, 1573(1995).
[7] R.Fehrenbacher and M.R. Norman, Phys. Rev. Lett. **74**, 3884 (1995).
[8] see e.g.: T.P. Devereaux, A. Virosztek and A. Zawadowski, Phy. Rev. B **51**, 505 (1995).
[9] E. Sherman, R. Li, and R. Feile, Solid State Commun., (in press).
[10] see e.g., Th. Kittel, *Quantum Theory of Solids* (Wiley, New York, 1963).
[11] K. Gofron, J.C. Campuzano, A.A. Abrikosov, M. Lindroos, A. Bansil, H. Ding, D. Koelling, and B. Dabrowski, Phys. Rev. Lett. **73**, 3302 (1994).
[12] W.-S. Zeng, W.Z. Lin, D. Mo, F.P. Pi, Z.J. Xia, Y.H. Zhou, G.C. Xiong, Z.Phys. B: Condensed Matter, **96**, 5 (1994).
[13] J.R. Schrieffer, *Theory of Superconductivity* (Addison-Wesley Publishing Company, Inc. 1988).
[14] C.C. Tsuei, D.M. Newns, C.C. Chi and P.C. Pattnaik, Phys. Rev. Lett. **65**, 2724 (1990).
[15] E.Dagotto, A.Nazarenko, and M.Boninsegni, Phys. Rev. Lett. **73**, 728 (1994); E.Dagotto, A.Nazarenko and A.Moreo, Phys. Rev. Lett. **74**, 310 (1995).